\documentstyle[12pt,subeqn]{article}

\makeatletter

\newtheorem{th}{Theorem}

\newcommand{\beq}{\begin{equation}}
\newcommand{\eeq}{\end{equation}}
\newcommand{\bea}{\begin{eqnarray}}
\newcommand{\eea}{\end{eqnarray}}
\newcommand{\bseq}{\begin{subequations}}
\newcommand{\eseq}{\end{subequations}}

\newcommand{\vecvar}[1]{\mbox{\boldmath$#1$}}

\def\endceqno{\end{array}\end{equation}}

\def\hf{\frac{1}{2}}
\let\nn=\nonumber
\def\beann{\begin{eqnarray*}} \def\eeann{\end{eqnarray*}}

\let\a=\alpha   \let\de=\delta
 \let\z=\zeta

 \let\Ds=\displaystyle

 \def\eq{\hspace{-2mm}&=&\hspace{-2mm}}
 \def\espace{\hspace{-2mm}&&\hspace{-2mm}}

 \def\0{\over } \def\1{\vec }    \def\2{{1\over2}} \def\4{{1\over4}}
 \def\5{\bar }  \def\6{\partial } \def\7#1{{#1}\llap{/}}

 \def\<{\langle } \def\>{\rangle }

 \let\ti=\tilde

 \def\i{{\rm i}} \def\tr{\mbox{tr}}

 \def\e{{\rm e}}

\begin{document}

\vspace*{20mm}
\noindent
{\LARGE New integrable systems of derivative nonlinear 
Schr\"{o}dinger equations with multiple components} 


\begin{quote}
Takayuki Tsuchida~\footnote{Electronic mail:
 tsuchida@monet.phys.s.u-tokyo.ac.jp} and Miki Wadati 

\bigskip
\noindent
{\it Department of Physics, Graduate School of Science, University of Tokyo,} 

\noindent
{\it  Hongo 7--3--1, Bunkyo-ku, Tokyo 113--0033, Japan} 

\bigskip
\noindent
(Received 25 June 1998, revised version received 05 January 1999)
\end{quote}

\noindent
The Lax pair for a derivative nonlinear Schr\"{o}dinger equation
proposed by Chen-Lee-Liu is generalized into matrix form. 
This gives new types of integrable coupled
derivative nonlinear Schr\"{o}dinger 
equations. By virtue of a gauge transformation, 
a new multi-component extension of a derivative nonlinear
Schr\"{o}dinger equation proposed by Kaup-Newell is also obtained. 


\bigskip
\noindent
keywords: derivative nonlinear Schr\"{o}dinger equation,
multi-component system, Lax pair, AKNS formulation, 
gauge transformation, conservation laws

\newpage
\section{Introduction}
\setcounter{equation}{0}
One of the most remarkable discoveries in soliton theory is the
inverse scattering method (ISM). It is well-known that there are two
types of derivative nonlinear Schr\"{o}dinger (DNLS) equation 
whose complete integrability can be proved by the ISM. One type
was proposed by Kaup and Newell \cite{KN},
\beq
\begin{array}{c}
\Ds \i \phi_t + \phi_{xx_{\vphantom M}} - \i (\phi \psi \phi)_x = 0, 
\\
\Ds \i \psi_t - \psi_{xx} - \i (\psi \phi \psi)_x = 0.
\end{array}
\label{}
\eeq
In what follows, we call this system the Kaup-Newell equation
 for convenience. The other type,
\beq
\begin{array}{c}
\Ds \i q_t + q_{xx_{\vphantom M}} - \i qr q_x = 0, 
\\
\Ds \i r_t - r_{xx} - \i rq r_x = 0,
\end{array}
\label{}
\eeq
was studied by Chen, Lee and Liu \cite{CLL}, 
who gave its recursion operator. We call this system the Chen-Lee-Liu
 equation. There is a simple transformation of dependent variables 
which cast one
into the other \cite{WS,Kund}. In this sense, these two types of the 
DNLS equation are gauge equivalent.

The ISM provides us a powerful tool to find multi-component extensions
of one-component soliton equations
\cite{AC} such as the nonlinear Schr\"{o}dinger (NLS) equation
 \cite{Manakov,Fordy1}, the modified KdV equation \cite{YO,Tsuchida1}, 
the Ablowitz-Ladik system \cite{Tsuchida2,Tsuchida3}. For the
Kaup-Newell equation, 
it has been known that 
there is a simple vector generalization \cite{Morris,Fordy2},
\beq
\i \vecvar{q}_t + \vecvar{q}_{xx} \mp \i (|\vecvar{q}|^2 \vecvar{q})_x 
= \vecvar{0}, 
\label{vKN}
\eeq
which are completely integrable. Besides the system (\ref{vKN}), 
Fordy \cite{Fordy2}
investigated various coupled versions of the Kaup-Newell equation by
considering Hermitian symmetric spaces. Yajima studied 
a generalization of coupled DNLS equations by means of a gauge
transformation in \cite{Yajima}. Meanwhile, multi-component
extensions of the Chen-Lee-Liu equation have not been studied
thoroughly from the ISM point of view. 

In recent years, multi-field
extensions or matrix generalizations of one-component classical
integrable systems have been developed considerably by means of
various approaches. Svinolupov, Sokolov, Habibullin and Yamilov
clarified close connections between soliton equations and Jordan
algebras or Jordan triple systems \cite{Svi1,Svi2,Svi3,HSY}. An 
excellent feature of their theory lies in the point that their approach
exhausts all integrable cases in some classes of multi-field
equations. Olver and Sokolov \cite{Olver} surveyed integrable systems 
whose dependent variables take their value in an associative algebra, 
{\it e.g.}, matrix-valued systems. They listed some classes of 
evolution equations on associative algebras which have higher-order 
symmetries.

In this Letter, we introduce a novel Lax formulation to get a 
matrix generalization of the Chen-Lee-Liu equation. As reductions of
the matrix equation, we obtain 
two types of coupled Chen-Lee-Liu equations. 
Through a transformation of variables, one type is cast into the
vector Kaup-Newell system (\ref{vKN}). It is noteworthy that the other
type is transformed into a new type of coupled Kaup-Newell
equations. The latter type of the coupled Chen-Lee-Liu equations is
shown to be connected with the coupled NLS equations,
%
%
\beq
\begin{array}{l}
\Ds \i u_{j,t} + u_{j,xx}
- 2\sum_{k=1_{\vphantom M}}^m u_k v_k \cdot u_j =0,
\\
\Ds \i v_{j,t} - v_{j,xx} + 2\sum_{k=1}^m v_k u_k \cdot v_j =0,
\end{array}
\hspace{10mm}
j=1,2, \cdots, m .
\label{CNLS}
\eeq
%
%
This indicates that the new coupled Chen-Lee-Liu equations and
the new coupled Kaup-Newell equations are solvable by the ISM.

The present Letter consists of the following. In section 2, we
introduce a new matrix generalization of the Lax pair for the
Chen-Lee-Liu equation. Under some reductions, we obtain two types of
coupled Chen-Lee-Liu equations. In addition, a systematic method to construct
conservation laws is given. In section 3, we perform gauge 
transformations to the coupled Chen-Lee-Liu equations
 found in the previous section. 
A connection between the coupled Chen-Lee-Liu equations and the coupled
NLS equations is clarified. In section 4, we propose 
transformations of variables to consider generalizations of the coupled 
Chen-Lee-Liu equations. With a particular choice of parameters, 
we obtain a new type of coupled Kaup-Newell equations. The
last section, section 5, is devoted to the concluding remarks. 

\section{Coupled DNLS equations}
\setcounter{equation}{0}
In this section, we consider a matrix generalization of 
the Chen-Lee-Liu equation by use of the ISM formulation. 
As reductions, we derive
two new integrable systems of coupled Chen-Lee-Liu equations.

\subsection{Lax formulation}
Let us begin with a system of linear equations,
\beq
\Psi_x = U \Psi, \hspace{6mm} \Psi_t = V \Psi.
\label{scattering}
\eeq
The compatibility condition $\Psi_{xt}=\Psi_{tx}$ is satisfied if
\beq
U_t -V_x +UV-VU = O.
\label{Lax_eq}
\eeq
We call $U$, $V$ the Lax pair and (\ref{Lax_eq}) the Lax equation 
or the zero-curvature condition. We choose the following form 
of the Lax pair,
\bea
U \eq
\i \z^2 \left[
\begin{array}{cc}
 -I_1  &  \\
    &  I_2 \\
\end{array}
\right]
+ 
\z \left[
\begin{array}{cc}
   &  Q \\
 R  &   \\
\end{array}
\right]
+ \i
\left[
\begin{array}{cc}
  O &  \\
    & \frac{1}{2}RQ  \\
\end{array}
\right],
\label{U_form}
\\
\vspace{-10mm}
\nn 
\\
V \eq 
\i \z^4 
\left[
\begin{array}{cc}
 -2I_1 &  \\
   & 2I_2  \\
\end{array}
\right]
+\z^3
\left[
\begin{array}{cc}
  & 2Q \\
 2R &  \\
\end{array}
\right]
+\i \z^2
\left[
\begin{array}{cc}
-QR  &  \\
  & RQ \\
\end{array}
\right]
\nn \\
\espace
 +\z
\left[
\begin{array}{cc}
  & \i Q_x + \frac{1}{2} QRQ \\
 -\i R_x + \frac{1}{2}RQR &  \\
\end{array}
\right]
+\i
\left[
\begin{array}{cc}
 O &  \\
  & \frac{\i}{2}(R Q_x - R_x Q) + \frac{1}{4}RQRQ \\
\end{array}
\right] .
\nn \\
&&
\label{V_form}
\eea
Here, $\z$ is the spectral parameter. $I_1$ and $I_2$ are respectively 
the $p \times p$ and the $q \times q$ identity matrices. $Q$ is a $p \times
q$ matrix and $R$ is a $q \times p$ matrix. Substituting (\ref{U_form})
and (\ref{V_form}) into (\ref{Lax_eq}) and equating the terms with the 
same powers of $\z$, we get a set of nonlinear evolution equations,
\begin{subequations}
\bea
&& \i Q_t + Q_{xx} -\i QRQ_x = O,
\\
&& \i R_t - R_{xx} -\i R_x QR = O.
\label{}
\eea
\label{matrix_eq}
\end{subequations}
Comments are in order. First, the equation obtained in 
the order $O(\z^0=1)$ of (\ref{Lax_eq})
 is automatically satisfied because of (\ref{matrix_eq}). Thus, we have 
 no restrictions on the sizes of $Q$ and $R$, that is, on $p$ and $q$.
This fact enables us to consider various multi-field extensions 
of the Chen-Lee-Liu equation by choosing the forms of $Q$ and 
$R$ appropriately. 
Second, the trace of $U$, $\tr \, U$, depends on dynamical variables
 in this formulation. Third, Olver and Sokolov showed that the matrix
 equation (\ref{matrix_eq}) possess at least one higher symmetry
 \cite{Olver}, which leads to a conjecture for the integrability of
 (\ref{matrix_eq}). The existence of the Lax pair gives a definite
 support to the complete integrability of the model.

\subsection{Coupled Chen-Lee-Liu equations (type I)}
As a reduction of (\ref{matrix_eq}), we choose $Q$ and $R$ to be a row
vector and a column vector respectively,
\beq
Q = (q_1, q_2, \cdots, q_m),
\hspace{5mm}
R = (r_1, r_2, \cdots, r_m)^T.
\label{}
\eeq
Here the superscript 
$T$ stands for the transposition. Then, we obtain a coupled 
version of the Chen-Lee-Liu equation,
\beq
\begin{array}{c}
\Ds 
\i q_{j,t}+q_{j,xx} - \i \sum_{k=1_{\vphantom M}}^m q_k r_k \cdot q_{j,x} = 0,
\nn \\
\Ds
\i r_{j,t}-r_{j,xx} - \i \sum_{k=1}^m r_k q_k \cdot r_{j,x} = 0,
\end{array}
\hspace{5mm} j=1, 2, \cdots, m.
\label{dnls1}
\eeq
In the following, we call this system the coupled Chen-Lee-Liu I
equations or, simply, the type I equations. 
The explicit form of the Lax matrix $U$ is given by
\beq
U=
\left[
\begin{array}{ccccc}
 -\i \z^2  & \z q_1 & \z q_2 & \cdots & \z q_m \\
 \z r_1  & \i \z^2 + \i \hf r_1 q_1 & \i \hf r_1 q_2 & 
\cdots & \i \hf r_1 q_m \\
 \z r_2 & \i \hf r_2 q_1 & \i \z^2+\i \hf r_2 q_2 & 
\cdots & \i \hf r_2 q_m  \\
 \vdots & \vdots & \vdots & \ddots & \vdots \\
 \z r_m & \i \hf r_m q_1 & \i \hf r_m q_2 & \cdots 
& \i \z^2+\i \hf r_m q_m \\
\end{array}
\right] .
\label{U_I}
\eeq
Under the reduction,
\beq
r_j = \pm q_j^{\ast}, \hspace{5mm} j=1, 2, \cdots, m,
\label{}
\eeq
the system (\ref{dnls1}) is expressed in a compact form,
\beq
\i \vecvar{q}_t + \vecvar{q}_{xx} \mp \i |\vecvar{q}|^2 \vecvar{q}_x 
= \vecvar{0},
\label{}
\eeq
with $\vecvar{q}$ being the vector,
\beq
\vecvar{q} = (q_1, q_2, \cdots, q_m).
\label{}
\eeq

\subsection{Conservation laws (type I)}
We can construct an
infinite number of conservation laws from the zero-curvature
 condition and the Lax pair. If we set
\beq
\Psi = (v_1, v_2, \cdots, v_{m+1})^T,
\label{}
\eeq
we have 
\beq
\Bigl( \sum_{j=1}^{m+1} U_{ij} v_j v_i^{-1}\Bigr)_t =
\Bigl( \sum_{j=1}^{m+1} V_{ij} v_j v_i^{-1}\Bigr)_x ,
\label{cons1}
\eeq
and
\beq
(U_{jj}-U_{ii}) v_j v_i^{-1} = - \sum_{k(\neq j)} U_{jk} v_k
v_i^{-1} + (v_j v_i^{-1})_x + v_j v_i^{-1} 
\sum_{k (\neq i)} U_{ik} v_k v_i^{-1} ,
\label{cons2}
\eeq
by virtue of (\ref{scattering}) and (\ref{Lax_eq}). Introducing a new
set of variables $\{ \Gamma_j \}$ by
\beq
\Gamma_j \equiv v_{j+1} v_1^{-1}, \hspace{5mm}j=1, 2, \cdots, m,
\label{}
\eeq
we get from (\ref{cons1}) and (\ref{cons2}) with (\ref{U_I}),
\beq
\Bigl( \sum_{j=1}^m q_j \Gamma_j \Bigr)_t = 
\Bigl( \z^{-1} V_{11} + \z^{-1} \sum_{j=1}^m V_{1 j+1} \Gamma_j \Bigr)_x ,
\label{cons3}
\eeq
and
\beq
q_j \Gamma_j = -\frac{1}{2\i \z}q_j r_j 
+ \frac{2\i}{(2\i \z)^2}q_j \Gamma_{j,x}
+ \frac{q_j r_j}{(2\i \z)^2}\sum_{k=1}^m q_k \Gamma_k 
+ \frac{1}{2\i \z} q_j \Gamma_j \sum_{k=1}^m q_k \Gamma_k .
\label{cons4}
\eeq
(\ref{cons3}) shows that $\sum_j q_j \Gamma_j$ is a generating
function of conserved densities. We expand $q_j \Gamma_j$ as
\beq
q_j \Gamma_j \equiv \sum_{l=1}^\infty \frac{1}{(2\i \z
  )^{2l-1}}f_j^{(l)} ,
\label{}
\eeq
to get recursion formulas for the conserved densities,
\beq
f_j^{(l)} = - q_j r_j \de_{l,1} + 2\i q_j 
(q_j^{-1}f_j^{(l-1)})_x + \sum_{n=2}^{l-1} f_j^{(n)} \sum_{k=1}^m 
f_k^{(l-n)} .
\label{cons5}
\eeq
For example, the followings are the first five conserved
densities for the coupled Chen-Lee-Liu I equations (\ref{dnls1}),
\bseq
\bea
I_1 \eq \sum_j q_j r_j,
\\
I_2 \eq q_j r_{l,x}, \hspace{3mm} \forall \; j,l,
{\vphantom \sum_{\vphantom M}}
\label{I_2}
 \\
I_3 \eq -4 \sum_j q_{j,x}r_{j,x} + \i \sum_j (q_j r_{j,x}-q_{j,x}r_j)
\cdot \sum_k q_k r_k,
\\
I_4 \eq -4 \i \sum_j (q_{j,x}r_{j,xx}-q_{j,xx}r_{j,x})
        +8\sum_j q_{j,x}r_{j,x} \sum_k q_k r_k
        -\Bigl\{ \sum_j (q_j r_{j,x}-q_{j,x}r_j) \Bigr\}^2 
+ \Bigl\{ \Bigl( \sum_j q_j r_j \Bigr)_x \Bigr\}^2
\nn \\
\espace  
       -\i \sum_j (q_j r_{j,x}-q_{j,x}r_j) \cdot 
      \Bigl( \sum_k q_k r_k \Bigr)^2,
\\
I_5 \eq -16 \sum_j q_{j,xx}r_{j,xx} +12\i \sum_j 
       (q_{j,x}r_{j,xx}-q_{j,xx}r_{j,x}) \cdot \sum_k q_k r_k
-4\i \sum_j (q_j r_{j,x}-q_{j,x}r_j)\cdot \Bigl(\sum_k q_k r_k
      \Bigr)_{xx} 
\nn \\
\espace      
   + 12\i \sum_j q_{j,x}r_{j,x} \cdot \sum_k 
      (q_k r_{k,x}-q_{k,x}r_k)
      -12\sum_j q_{j,x}r_{j,x}\cdot 
      \Bigl(\sum_k q_k r_k \Bigr)^2 -3 \sum_j q_j r_j \cdot 
      \Bigl\{ \Bigl( \sum_k q_k r_k \Bigr)_x \Bigr\}^2 
\nn \\
\espace    + 3\Bigl\{ 
      \sum_j (q_j r_{j,x}-q_{j,x}r_j)\Bigr\}^2 \cdot \sum_k q_k r_k
      + \i \sum_j (q_j r_{j,x}-q_{j,x}r_j)\cdot 
      \Bigl(\sum_k q_k r_k \Bigr)^3.
\label{}
\eea
\eseq
%

\subsection{Coupled Chen-Lee-Liu equations (type II)}
As another reduction of (\ref{matrix_eq}), we choose $Q$ and $R$ to be a 
column vector and a row vector respectively,
\beq
Q = (\ti{q}_1, \ti{q}_2, \cdots, \ti{q}_m)^T,
\hspace{5mm}
R = (\ti{r}_1, \ti{r}_2, \cdots, \ti{r}_m).
\label{}
\eeq
In this case, (\ref{matrix_eq}) reduces to
\beq
\begin{array}{c}
\Ds
\i \ti{q}_{j,t}+\ti{q}_{j,xx} - \i \sum_{k=1_{\vphantom M}}^m
\ti{q}_{k,x} \ti{r}_k \cdot \ti{q}_{j} = 0,
\\
\Ds
\i \ti{r}_{j,t}-\ti{r}_{j,xx} - \i \sum_{k=1}^m \ti{r}_{k,x} \ti{q}_k 
\cdot \ti{r}_{j} = 0,
\end{array}
\hspace{5mm} j=1, 2, \cdots, m.
\label{dnls2}
\eeq
We call this system the coupled Chen-Lee-Liu II equations or, 
simply, the type II equations in what follows. 
Hisakado proposed the coupled Chen-Lee-Liu II equations 
independently in \cite{Hisakado1}. 
The Lax matrix $U$ for (\ref{dnls2}) is given by
\beq
U=
\left[
\begin{array}{cccc}
 -\i \z^2  & & & \z \ti{q}_1 \\
  & \ddots & & \vdots \\
  & & -\i \z^2 & \z \ti{q}_m \\
 \z \ti{r}_1 & \cdots & \z \ti{r}_m 
& \i \z^2 +\i \hf \sum_{k=1}^m \ti{r}_k \ti{q}_k \\
\end{array}
\right] .
\label{U_II}
\eeq
The representation of the Lax pair is not unique. We can employ
a different representation of $U$ for (\ref{dnls2}) as
\beq
U=
\left[
\begin{array}{cccc}
 -\i \z^2 -\i \hf \sum_{k=1}^m \ti{q}_k \ti{r}_k & \z \ti{q}_1 & \cdots 
& \z \ti{q}_m \\
 \z \ti{r}_1 & \i \z^2 & &  \\
 \vdots & & \ddots & \\
 \z \ti{r}_m & & & \i \z^2 \\
\end{array}
\right] .
\label{U_II'}
\eeq
The difference in the structure of the Lax matrices (\ref{U_I})
 and (\ref{U_II}) (or (\ref{U_II'})) 
should be noteworthy. We can calculate an infinite
 set of conservation laws for (\ref{dnls2}) in the same manner as in 
section 2.3. The recursion relations for the conserved densities are
\beq
g_j^{(l)} = -q_j r_j \de_{l,1} + 2\i q_j (q_j^{-1}g_j^{(l-1)})_x 
        + \sum_{n=1}^{l-2} g_j^{(n)}\sum_{k=1}^m g_k^{(l-n)},
\label{}
\eeq
which yield an infinite series of conserved densities $\sum_l g_j^{(l)}$.
The first five conserved densities are, for instance,
\bseq
\bea
\ti{I}_1 \eq \sum_j \ti{q}_j \ti{r}_j,
\\
\ti{I}_2 \eq \i \sum_j (\ti{q}_j \ti{r}_{j,x}-\ti{q}_{j,x}\ti{r}_j), 
{\vphantom \sum_{\vphantom M}}
\label{ti_I_2}
\\
\ti{I}_3 \eq -4 \sum_j \ti{q}_{j,x}\ti{r}_{j,x} + \i \sum_j 
(\ti{q}_j \ti{r}_{j,x}-\ti{q}_{j,x}\ti{r}_j)
\cdot \sum_k \ti{q}_k \ti{r}_k,
\\
\ti{I}_4 \eq -4 \i \sum_j (\ti{q}_{j,x}\ti{r}_{j,xx}-\ti{q}_{j,xx}\ti{r}_{j,x})
        +4\sum_j \ti{q}_{j,x}\ti{r}_{j,x} \sum_k \ti{q}_k \ti{r}_k
        -2\Bigl\{ \sum_j (\ti{q}_j \ti{r}_{j,x}-\ti{q}_{j,x}\ti{r}_j) 
        \Bigr\}^2 +2 \Bigl\{ \Bigl( \sum_j \ti{q}_j \ti{r}_j \Bigr)_x \Bigr\}^2
\nn \\
\espace  
       -\i \sum_j (\ti{q}_j \ti{r}_{j,x}-\ti{q}_{j,x}\ti{r}_j) \cdot 
      \Bigl( \sum_k \ti{q}_k \ti{r}_k \Bigr)^2,
\\
\ti{I}_5 \eq -16 \sum_j \ti{q}_{j,xx}\ti{r}_{j,xx} +4\i \sum_j 
       (\ti{q}_{j,x}\ti{r}_{j,xx}-\ti{q}_{j,xx}\ti{r}_{j,x}) 
       \cdot \sum_k \ti{q}_k \ti{r}_k
       -12\i \sum_j (\ti{q}_j \ti{r}_{j,x}-\ti{q}_{j,x}\ti{r}_j)
       \cdot \Bigl(\sum_k \ti{q}_k \ti{r}_k \Bigr)_{xx} 
\nn \\
\espace      
   + 20\i \sum_j \ti{q}_{j,x}\ti{r}_{j,x} \cdot \sum_k 
      (\ti{q}_k \ti{r}_{k,x}-\ti{q}_{k,x}\ti{r}_k)
      -4\sum_j \ti{q}_{j,x}\ti{r}_{j,x}\cdot 
      \Bigl(\sum_k \ti{q}_k \ti{r}_k \Bigr)^2 -5 \sum_j \ti{q}_j
      \ti{r}_j \cdot \Bigl\{ \Bigl( \sum_k \ti{q}_k \ti{r}_k \Bigr)_x 
      \Bigr\}^2 
\nn \\
\espace    + 5\Bigl\{ 
      \sum_j (\ti{q}_j \ti{r}_{j,x}-\ti{q}_{j,x}\ti{r}_j)\Bigr\}^2 
      \cdot \sum_k \ti{q}_k \ti{r}_k
      + \i \sum_j (\ti{q}_j \ti{r}_{j,x}-\ti{q}_{j,x}\ti{r}_j)\cdot 
      \Bigl(\sum_k \ti{q}_k \ti{r}_k \Bigr)^3.
\label{}
\eea
\eseq
It is interesting to compare (\ref{I_2}) and (\ref{ti_I_2}): 
$\ti{I}_2$ is a conserved density only after 
taking a summation with respect to the subscript $j$. 
%
%

\section{Gauge transformations}
\setcounter{equation}{0}
In this section, we investigate the structure of the Lax pairs given
in the previous section in connection with the AKNS formulation
\cite{AKNS}. For this purpose, we introduce a gauge
transformation \cite{WS},
\beq
\Psi = g \Phi,
\label{}
\eeq
which changes the Lax matrices $U$, $V$ into
\bseq
\bea
\Ds U'_{\vphantom M} \eq g^{-1} U g - g^{-1} g_x ,
\\
\Ds V'^{\vphantom M} \eq g^{-1} V g - g^{-1} g_t .
\label{}
\eea
\eseq
In the following, dependent variables are assumed to approach $0$ as
$|x| \to \infty$ for convenience.
\subsection{type I}
In terms of $\{q_j\}$, $\{r_j\}$ which satisfy (\ref{dnls1}), 
we introduce a new set of variables $\{ u_j\}$, $\{v_j\}$ by
\beq
\begin{array}{l}
\Ds u_j = a q_j \exp\Bigl\{_{\vphantom M} -\frac{\i}{2} \int_{-\infty}^x 
\sum_{k=1_{\vphantom M}}^m q_k r_k {\rm d}x' \Bigr\},
\\
\Ds v_j = b r_{j,x} \exp \Bigl\{ \frac{\i}{2} \int_{-\infty}^x 
\sum_{k=1}^m q_k r_k {\rm d}x' \Bigr\},
\end{array}
\hspace{5mm} j=1, 2, \cdots, m.
\label{Miura}
\eeq
Here the constants $a$ and $b$ satisfy $ab = -\i/2$. Using 
(\ref{Miura}) and the first conservation law for (\ref{dnls1}), we obtain
\bseq
\beq
\i u_{j,t}
 + u_{j,xx}- 2\sum_{k=1_{\vphantom M}}^m u_k v_k \cdot u_j 
= \Bigl[ \i q_{j,t}+q_{j,xx}-\i \sum_{k=1}^m q_k r_k \cdot q_{j,x}
\Bigr] a \exp \Bigl\{ -\frac{\i}{2} \int_{-\infty}^x 
\sum_{k=1_{\vphantom M}}^m q_k r_k {\rm d}x' \Bigr\} ,
\label{}
\eeq
\beq
\i v_{j,t} - v_{j,xx}+ 2\sum_{k=1}^m v_k u_k \cdot v_j 
= \Bigl[ \i r_{j,t} -r_{j,xx}-\i \sum_{k=1}^m r_k q_k \cdot r_{j,x}
\Bigr]_x b \exp\Bigl\{ \frac{\i}{2} \int_{-\infty}^x 
\sum_{k=1_{\vphantom M}}^m q_k r_k {\rm d}x' \Bigr\} .
\label{}
\eeq
\eseq
Hence, we conclude that if $\{q_j \}$ and $\{ r_j \}$ satisfy the
coupled Chen-Lee-Liu I equations, $\{u_j \}$ and $\{v_j \}$ satisfy
the well-known coupled NLS equations,
\beq
\begin{array}{l}
\Ds \i u_{j,t}
 + u_{j,xx}- 2\sum_{k=1_{\vphantom M}}^m u_k v_k \cdot u_j =0,
\\
\Ds \i v_{j,t} - v_{j,xx}+ 2\sum_{k=1}^m v_k u_k \cdot v_j =0,
\end{array}
\hspace{10mm}
j=1,2, \cdots, m .
\label{CNLS2}
\eeq
By means of the transformation (\ref{Miura}), 
the second conserved densities (\ref{I_2}) are cast into the
second conserved densities for (\ref{CNLS2}),
\beq
I_2' = u_j v_{l}, \hspace{3mm} \forall \; j,l .
\label{}
\eeq
The Lax matrix for the coupled NLS equations (\ref{CNLS2}) is obtained 
from (\ref{U_I}) through a gauge transformation,
\beq
g=
\left[
\begin{array}{cccc}
 2\i b \z \e^{K} & 0 & \cdots & 0 \\
 -b r_1 \e^{K} & 1 &  &  \\
 \vdots &  & \ddots &  \\
 -b r_m \e^{K} &  & & 1 \\
\end{array}
\right],
\hspace{5mm} K=\frac{\i}{2}\int_{-\infty}^x 
\sum_{k=1_{\vphantom M}}^m q_k r_k {\rm d}x',
\label{}
\eeq
as
\bea
U' \eq
g^{-1} U g - g^{-1} g_x 
\nn \\
\eq
\left[
\begin{array}{cccc}
 -\i \z^2  & u_1 & \cdots & u_m \\
 v_1  & \i \z^2 &  &  \\
 \vdots &  & \ddots &  \\
 v_m &  & & \i \z^2  \\
\end{array}
\right] .
\label{}
\eea
We thus have shown that the scattering problem for the coupled
Chen-Lee-Liu I equations associated with (\ref{U_I}) is gauge
equivalent to $sl \, (m+1)$ AKNS formulation.
\subsection{type II}
Next, let us consider a gauge transformation of the Lax matrix for the 
coupled Chen-Lee-Liu II equations. By virtue of a gauge transformation,
\beq
g
=
\left[
\begin{array}{cccc}
 2\i b \z  &  &  &  \\
  & \ddots &  &  \\
  &  & 2\i b \z  &  \\
 -b \ti{r}_1 & \cdots & -b \ti{r}_m & 1 \\
\end{array}
\right] ,
\label{}
\eeq
the Lax matrix (\ref{U_II}) is cast into 
%
\bea
U'
\eq
g^{-1} U g - g^{-1} g_x
\nn \\
\eq
\left[
\begin{array}{cccc}
 -\i \z^2 &  &  &  \\
  & \ddots &  &  \\
  &  & -\i \z^2 &  \\
  &  &  & \i \z^2 \\
\end{array}
\right]
+
\left[
\begin{array}{cccc}
 \i \hf \ti{q}_1 \ti{r}_1 & \cdots & \i \hf \ti{q}_1 \ti{r}_m & a \ti{q}_1 \\
 \vdots & \ddots & \vdots & \vdots \\
 \i \hf \ti{q}_m \ti{r}_1 & \cdots & \i \hf \ti{q}_m \ti{r}_m & a \ti{q}_m \\
 b \ti{r}_{1,x} & \cdots & b \ti{r}_{m,x} & 0 \\
\end{array}
\right] ,
\label{}
\eea
with $ab = -\i/2$. 
It is noticed that there is no longer the spectral parameter $\z$ in
the potential part of $U'$. We can easily eliminate dependent 
variables in the diagonal
elements of $U'$ by a further transformation. Thus, the
gauge-transformed Lax
formulation is embedded in 
the $(m+1) \times (m+1)$ matrix generalization of the AKNS formulation 
\cite{ZS2,AH}.

\section{Generalizations}
\setcounter{equation}{0}
In this section, we consider generalizations of the 
coupled Chen-Lee-Liu equations (type I $\&$ II) via transformations of 
variables. 
\subsection{type I}
\label{type_I}
We first transform the coupled Chen-Lee-Liu I equations into a system 
with cubic terms without differentiation. Let us change independent
and dependent variables by
\beq
T=t, \hspace{5mm} X = x + \frac{2\beta}{\a} t,\hspace{20mm}
 \label{}
\eeq
\beq
\begin{array}{l}
\Ds q_j = \sqrt{\a} Q_j \exp \Bigl\{-\i \frac{\beta}{\a}X 
+ \i \Bigl(\frac{\beta}{\a_{\vphantom M}}\Bigr)^2 T \Bigr\}_{\vphantom M},
\\
\Ds r_j = \sqrt{\a} R_j \exp \Bigl\{\i \frac{\beta}{\a}X 
- \i \Bigl(\frac{\beta^{\vphantom M}}{\a}\Bigr)^2 T \Bigr\}^{\vphantom M},
\end{array}
\hspace{5mm} j=1,2, \cdots,m.
\label{}
\eeq
From (\ref{dnls1}), we get a system of equations,
\beq
\begin{array}{c}
\Ds 
\i Q_{j,T}+Q_{j,XX} - \i \a \sum_{k=1_{\vphantom M}}^m Q_k R_k 
\cdot Q_{j,X} - \beta \sum_{k=1}^m Q_k R_k \cdot Q_j = 0,
\nn \\
\Ds
\i R_{j,T}-R_{j,XX}  - \i \a \sum_{k=1}^m R_k Q_k \cdot R_{j,X} 
+ \beta \sum_{k=1}^m R_k Q_k \cdot R_j = 0,
\end{array}
\hspace{5mm} j=1, 2, \cdots, m.
\label{}
\eeq
We further utilize a kind of gauge transformation,
\beq
\begin{array}{l}
\Ds \phi_j = Q_j \exp\Bigl\{_{\vphantom M} -2\i \de \int_{-\infty}^X 
\sum_{k=1_{\vphantom M}}^m Q_k R_k {\rm d}X' \Bigr\},
\\
\Ds \psi_j = R_j \exp \Bigl\{ 2\i \de \int_{-\infty}^X 
\sum_{k=1}^m Q_k R_k {\rm d}X' \Bigr\},
\end{array}
\hspace{5mm} j=1, 2, \cdots, m,
\label{}
\eeq
and get an extension of (\ref{dnls1}),
\beq
\begin{array}{l}
\Ds \i \phi_{j,T} + \phi_{j,XX} -\beta 
\sum_{k=1_{\vphantom M}}^m \phi_k \psi_k \cdot \phi_j 
+4\i \de \sum_{k=1}^m \phi_k \psi_{k,X} \cdot \phi_j 
\\
\Ds + \i (4\de -\a)\sum_{k=1}^m \phi_k \psi_k \cdot \phi_{j,X}
+ \de (4\de + \a) \Bigl( \sum_{k=1}^m \phi_k \psi_k \Bigr)^2 \phi_j
=0,
\\
\Ds \i \psi_{j,T} - \psi_{j,XX} +\beta 
\sum_{k=1}^m \psi_k \phi_k \cdot \psi_j 
+4\i \de \sum_{k=1}^m \psi_k \phi_{k,X} \cdot \psi_j 
\\
\Ds + \i (4\de-\a)\sum_{k=1}^m \psi_k \phi_k \cdot \psi_{j,X}
-\de (4\de+\a) \Bigl( \sum_{k=1}^m \psi_k \phi_k \Bigr)^2 \psi_j 
=0,
\end{array}
\hspace{5mm} j=1, 2, \cdots, m.
\label{}
\eeq
For a choice of $4\de + \a=0$, the system reads
\beq
\begin{array}{l}
\Ds \i \phi_{j,T} + \phi_{j,XX}
-\i \a \sum_{k=1}^m \phi_k \psi_{k,X} \cdot \phi_j 
-2\i \a \sum_{k=1}^m \phi_k \psi_k \cdot \phi_{j,X}
-\beta \sum_{k=1_{\vphantom M}}^m \phi_k \psi_k \cdot \phi_j 
=0,
\\
\Ds \i \psi_{j,T} - \psi_{j,XX}
- \i \a \sum_{k=1}^m \psi_k \phi_{k,X} \cdot \psi_j 
 -2\i \a \sum_{k=1}^m \psi_k \phi_k \cdot \psi_{j,X}
+\beta \sum_{k=1}^m \psi_k \phi_k \cdot \psi_j 
=0,
\end{array}
\hspace{5mm} j=1, 2, \cdots, m.
\label{cKN1}
\eeq
For $m=1$, this is equivalent to the Kaup-Newell equation with a cubic
term \cite{WKI}. Thus, the system (\ref{cKN1}) 
is interpreted as a new multi-field extension of the Kaup-Newell
equation. We can give the explicit expression of the corresponding Lax
pair. For instance, the Lax matrix for (\ref{cKN1}) with $\a=1$,
$\beta=0$ is given by
\beq
U=
\left[
\begin{array}{ccccc}
 -\i \z^2 + \i \hf \sum_{k=1}^m \phi_k \psi_k & \z \phi_1 & \z \phi_2 & 
\cdots & \z \phi_m \\
 \z \psi_1  & \i \z^2 + \i \hf \psi_1 \phi_1 & \i \hf \psi_1 \phi_2 & 
\cdots & \i \hf \psi_1 \phi_m \\
 \z \psi_2 & \i \hf \psi_2 \phi_1 & \i \z^2+\i \hf \psi_2 \phi_2 & 
\cdots & \i \hf \psi_2 \phi_m  \\
 \vdots & \vdots & \vdots & \ddots & \vdots \\
 \z \psi_m & \i \hf \psi_m \phi_1 & \i \hf \psi_m \phi_2 & \cdots 
& \i \z^2+\i \hf \psi_m \phi_m \\
\end{array}
\right] .
\label{}
\eeq
\subsection{type II}
Following the same procedure as type I, we obtain a
generalization of the coupled Chen-Lee-Liu II equations
 (\ref{dnls2}). By a change of variables,
\beq
T=t, \hspace{5mm} X = x + \frac{2\beta}{\a} t, \hspace{20mm}
\label{}
\eeq
\beq
\begin{array}{l}
\Ds \ti{q}_j = \sqrt{\a} \ti{Q}_j \exp \Bigl\{-\i \frac{\beta}{\a}X 
+ \i \Bigl(\frac{\beta}{\a_{\vphantom M}}\Bigr)^2 T \Bigr\}_{\vphantom M},
\\
\Ds \ti{r}_j = \sqrt{\a} \ti{R}_j \exp \Bigl\{\i \frac{\beta}{\a}X 
- \i \Bigl(\frac{\beta^{\vphantom M}}{\a}\Bigr)^2 T \Bigr\}^{\vphantom M},
\end{array}
\hspace{5mm} j=1,2, \cdots,m ,
\label{}
\eeq
we get a system of equations,
\beq
\begin{array}{c}
\Ds 
\i \ti{Q}_{j,T}+\ti{Q}_{j,XX} - \i \a \sum_{k=1_{\vphantom M}}^m
\ti{Q}_{k,X} \ti{R}_k 
\cdot \ti{Q}_{j} - \beta \sum_{k=1}^m \ti{Q}_k \ti{R}_k \cdot \ti{Q}_j = 0,
\nn \\
\Ds
\i \ti{R}_{j,T}-\ti{R}_{j,XX}  - \i \a \sum_{k=1}^m \ti{R}_{k,X}
\ti{Q}_k \cdot \ti{R}_{j} 
+ \beta \sum_{k=1}^m \ti{R}_k \ti{Q}_k \cdot \ti{R}_j = 0,
\end{array}
\hspace{5mm} j=1, 2, \cdots, m.
\label{}
\eeq
By virtue of a gauge transformation,
\beq
\begin{array}{l}
\Ds \ti{\phi}_j = \ti{Q}_j \exp\Bigl\{_{\vphantom M} 
-2\i \de \int_{-\infty}^X 
\sum_{k=1_{\vphantom M}}^m \ti{Q}_k \ti{R}_k {\rm d}X' \Bigr\},
\\
\Ds \ti{\psi}_j = \ti{R}_j \exp \Bigl\{ 2\i \de \int_{-\infty}^X 
\sum_{k=1}^m \ti{Q}_k \ti{R}_k {\rm d}X' \Bigr\},
\end{array}
\hspace{5mm} j=1, 2, \cdots, m,
\label{}
\eeq
we obtain
\beq
\begin{array}{l}
\Ds \i \ti{\phi}_{j,T} + \ti{\phi}_{j,XX} -\beta 
\sum_{k=1_{\vphantom M}}^m \ti{\phi}_k \ti{\psi}_k \cdot \ti{\phi}_j 
- \i \a \sum_{k=1}^m \ti{\phi}_{k,X} \ti{\psi}_k \cdot \ti{\phi}_{j}
+4\i \de \sum_{k=1}^m \ti{\phi}_k \ti{\psi}_{k,X} \cdot \ti{\phi}_j 
\\
\Ds + 4\i \de \sum_{k=1}^m \ti{\phi}_k \ti{\psi}_k \cdot \ti{\phi}_{j,X}
+ \de (4\de + \a) \Bigl( \sum_{k=1}^m \ti{\phi}_k \ti{\psi}_k
\Bigr)^2 \ti{\phi}_j =0,
\\
\Ds \i \ti{\psi}_{j,T} - \ti{\psi}_{j,XX} +\beta 
\sum_{k=1}^m \ti{\psi}_k \ti{\phi}_k \cdot \ti{\psi}_j 
- \i \a \sum_{k=1}^m \ti{\psi}_{k,X} \ti{\phi}_k \cdot \ti{\psi}_{j}
+4\i \de \sum_{k=1}^m \ti{\psi}_k \ti{\phi}_{k,X} \cdot \ti{\psi}_j 
\\
\Ds + 4\i \de \sum_{k=1}^m \ti{\psi}_k \ti{\phi}_k \cdot \ti{\psi}_{j,X}
-\de (4\de+\a) \Bigl( \sum_{k=1}^m \ti{\psi}_k \ti{\phi}_k \Bigr)^2 
\ti{\psi}_j =0,
\end{array}
\hspace{5mm} j=1, 2, \cdots, m.
\label{}
\eeq
In case of $4\de + \a=0$, the system essentially reduces to 
the well-known vector Kaup-Newell system (\ref{vKN}) in a slightly
different notation,
\beq
\begin{array}{l}
\Ds \i \ti{\phi}_{j,T} + \ti{\phi}_{j,XX}
-\i \a \Bigl( \sum_{k=1}^m \ti{\phi}_k \ti{\psi}_{k} \cdot \ti{\phi}_j 
\Bigr)_X 
-\beta \sum_{k=1_{\vphantom M}}^m \ti{\phi}_k \ti{\psi}_k \cdot \ti{\phi}_j =0,
\\
\Ds \i \ti{\psi}_{j,T} - \ti{\psi}_{j,XX}
- \i \a \Bigl( \sum_{k=1}^m \ti{\psi}_k \ti{\phi}_{k} \cdot
\ti{\psi}_j \Bigr)_X
+ \beta \sum_{k=1}^m \ti{\psi}_k \ti{\phi}_k \cdot \ti{\psi}_j =0,
\end{array}
\hspace{5mm} j=1, 2, \cdots, m.
\label{cKN2}
\eeq
This shows that the coupled Chen-Lee-Liu II equations are gauge
equivalent to the vector Kaup-Newell system. 
The Lax matrix for (\ref{cKN2}) with $\a=1$, $\beta=0$ is given by
\cite{Fordy1,Morris}
\beq
U=
\left[
\begin{array}{cccc}
 -\i \z^2 &  &  & \z \ti{\phi}_1 \\
  & \ddots &  &  \vdots \\
   &  & -\i \z^2 & \z \ti{\phi}_m \\
 \z \ti{\psi}_1 & \cdots & \z \ti{\psi}_m  & \i \z^2 \\
\end{array}
\right] .
\label{}
\eeq
This can also be derived from (\ref{U_II}) by means of a gauge transformation.

\section{Concluding remarks}
\setcounter{equation}{0}
In this Letter, we have presented a matrix formulation of the 
inverse scattering method (ISM) and have found the two types of the coupled
Chen-Lee-Liu equations, (\ref{dnls1}) and (\ref{dnls2}). 
As is often the case with one-component soliton systems 
\cite{Fordy1,Fordy2,Svi1,Svi2,Svi3,HSY,Olver}, the 
Chen-Lee-Liu equation also has plural multi-field generalizations. 
Using the Lax pairs, the conservation laws and the
gauge transformations, we have studied the properties of the two
types of the coupled Chen-Lee-Liu equations. 
An important step to obtain the coupled derivative NLS equations in our theory
is the introduction of $U$ by (\ref{U_form}). The form of $U$
seems unusual, because we more or less assume ${\rm tr}\, U$ to be a 
constant (often equal to $0$) for soliton systems. 
The Lax pair is,
 however, transformed into the well-known one for the coupled NLS equations 
($sl(m+1)\,$AKNS formulation), which is solvable by the ISM. 
The ISM for the coupled
Chen-Lee-Liu I equations (\ref{dnls1}) will be reported in detail in 
a separate paper. 

In the two-component case, the coupled Chen-Lee-Liu equations
 (type I \& II) and the related models maybe have physical
 significances. They may describe wave propagations in birefringent optical
 fibers with nonlinear effects such as the Raman scattering and the
 Kerr effect.

The coupled Chen-Lee-Liu equations reported in this Letter have an
infinite number of conservation laws and are completely integrable. We
can construct other flows of the hierarchies by employing corresponding
 time-dependences of the scattering problems ($\z$-dependences of the 
Lax matrix $V$). These flows have in common the conserved densities 
for the original flows. Expanding the matrix $V$ 
 from $O(\z^6)$ to $O(1)$, we get
\beq
\begin{array}{r}
\Ds
q_{j,t} + \frac{1}{2}q_{j,xxx}-\i \frac{3}{4}\Bigl(\sum_{k} q_{k,x}r_k
q_{j,x} + \sum_{k} q_k r_k q_{j,xx} \Bigr)
- \frac{3}{8} 
\Bigl( \sum_{k_{\vphantom |}} q_k r_k \Bigr)^2 q_{j,x} 
=0,
\\
\Ds
r_{j,t} + \frac{1}{2}r_{j,xxx}+\i \frac{3^{\vphantom L}}{4}
\Bigl(\sum_{k} r_{k,x}q_k
r_{j,x} + \sum_{k} r_k q_k r_{j,xx}\Bigr) 
- \frac{3}{8} \Bigl( \sum_{k} r_k q_k \Bigr)^2 r_{j,x} 
=0,
\end{array}
\label{}
\eeq
for the coupled Chen-Lee-Liu I hierarchy (cf. (\ref{U_I})), and
\beq
\begin{array}{r}
\Ds
\ti{q}_{j,t} + \hf \ti{q}_{j,xxx}-\i \frac{3}{4}\Bigl(\sum_{k} \ti{q}_{k,x}\ti{r}_k
\ti{q}_{j,x} + \sum_{k} \ti{q}_{k,xx} \ti{r}_k \ti{q}_{j}\Bigr)
- \frac{3}{8} 
\sum_{k} \ti{q}_k \ti{r}_k \sum_{l_{\vphantom |}} 
\ti{q}_{l,x} \ti{r}_l \ti{q}_{j} =0,
\\
\Ds
\ti{r}_{j,t} + \frac{1}{2}\ti{r}_{j,xxx}+\i \frac{3^{\vphantom
    L}}{4}\Bigl(\sum_{k} \ti{r}_{k,x} \ti{q}_k
\ti{r}_{j,x} + \sum_{k} \ti{r}_{k,xx} \ti{q}_k \ti{r}_j \Bigr) 
- \frac{3}{8} \sum_{k} \ti{r}_k \ti{q}_k \sum_{l} \ti{r}_{l,x}
 \ti{q}_l \ti{r}_j =0,
\end{array}
\label{}
\eeq
for the coupled Chen-Lee-Liu II hierarchy (cf. (\ref{U_II})). 

On the other hand, by expanding $V$ from $O(\z^{-2})$ to $O(1)$, we
obtain coupled versions of the massive Thirring model in a light-cone 
frame \cite{Tsuchida4},
\beq
\begin{array}{l}
\Ds
q_{j,t}-2\i \varphi_j -\i \hf \sum_k \varphi_k \sigma_k q_j =0,
\\
\Ds
r_{j,t}+2\i \sigma_j + \i \hf \sum_k \sigma_k \varphi_k r_j =0,
\\
\Ds
\varphi_{j,x} + 2\i q_j +\i \hf \sum_k \varphi_k r_k q_j = 0,
\\
\Ds
\sigma_{j,x} -2\i r_j -\i \hf \sum_k \sigma_k q_k r_j = 0,
\end{array}
\label{Thirring1}
\eeq
for the coupled Chen-Lee-Liu I hierarchy, and
\beq
\begin{array}{l}
\Ds
\ti{q}_{j,t}-2\i \ti{\varphi}_j -\i \hf \sum_k \ti{q}_k \ti{\sigma}_k 
\ti{\varphi}_j =0,
\\
\Ds
\ti{r}_{j,t}+2\i \ti{\sigma}_j + \i \hf \sum_k \ti{r}_k \ti{\varphi}_k
 \ti{\sigma}_j =0,
\\
\Ds
\ti{\varphi}_{j,x} + 2\i \ti{q}_j +\i \hf \sum_k \ti{q}_k \ti{r}_k 
\ti{\varphi}_j = 0,
\\
\Ds
\ti{\sigma}_{j,x} -2\i \ti{r}_j -\i \hf \sum_k \ti{r}_k \ti{q}_k 
\ti{\sigma}_j = 0,
\end{array}
\label{Thirring2}
\eeq
for the coupled Chen-Lee-Liu II hierarchy. Here, the explicit 
form of $V$ is expressed as
\beq
V = \i \z^{-2} 
\left[
\begin{array}{cc}
 I   &  \\
  & -I \\
\end{array}
\right]
+ \z^{-1}
\left[
\begin{array}{cc}
   &  \varphi  \\
 \sigma  &  \\
\end{array}
\right]
+ \i
\left[
\begin{array}{cc}
 \frac{1}{2} \varphi \sigma  &  \\
   & O \\
\end{array}
\right].
\eeq
$\varphi_j$ ($\ti{\varphi}_j$) and $\sigma_j$ ($\ti{\sigma}_j$) 
are components of a row (column) vector $\varphi$ 
and a column (row) vector $\sigma$ for the system (\ref{Thirring1}) 
((\ref{Thirring2})) respectively. 
By a simple change of variables such as $t \leftrightarrow x$, 
(\ref{Thirring1}) and (\ref{Thirring2}) are mutually transformed 
into the other. 
The split into two different systems is not
essential at the level of these flows. 
%
%


\end{document}